# Computer Based Interpretation of the Students' Evaluation of the Teaching Staff

**Tiberiu Marius Karnyanszky, Ovidiu Crista, Cătălin Ţuican**
„Tibiscus" University of Timişoara

ABSTRACT: The goal of this paper is to offer a full support for universities and quality assessment committees in retrieving the feedback from their students regarding to their teaching staff. The computer based application presented before ([Cri07]) collects data from the students. Another part of the application, presented in this paper, processes this data and presents the statistical results concerning each teacher.

**Introduction**

Based on the 222/2008 Project implemented by the Faculty of Computers and Applied Computer Science from the "Tibiscus" University, winner of the 2007 Tibiscus Research Competition, the *Online Evaluation* ([Ara06, Cri07, Dra09, TKS08]) is a computer based program that allows:
- For students, to evaluate their teachers;
- For teachers, to be evaluated and to obtain a statistical processed result of the evaluation;
- For quality assessment committee, to validate the evaluations and to perform a statistical processing of the valid results;
- For faculties/universities management, to obtain a feed-back from their students and to apply adequate measures to increase the quality of the educational process.

The evaluation in our university bases on a questionnaire using 58 questions regarding ([K+08]):

- The scientific competence (12 items);
- The psycho pedagogical competence (20 items);
- The psychosocial competence (13 items);
- The managerial competence (13 items).

**Collected Data**

The application uses a MySQL Database including a 'rezultate" table ([TKS08]) containing the students' answers to the questionnaire. Only the complete questionnaires (58 answers) are memorized into the table so only the valid records are included to further processing.

This database table contains an ID for the questionnaire (to maintain an order of the records), a time indicator (shows the moment of the evaluation), an ID of the evaluated teacher (all answers are included in the same database) and 58 marks (representing the students' answers: 1=very poor, 2=poor, 3=medium, 4=good, 5=very good).

Each student can see his own answers but cannot see others questionnaires. Only the systems' administrator can see the entire database and he also can ask the application to make the statistical processing for a specified teacher.

**Processed Data**

The 58 answers memorized into the database refer to four different competencies of the teacher. Let's start with the results for *Teacher-1* presented in Table 1, selected from the entire database using a filter conditioned by the teachers' ID:

**Table 1: Questionnaires' results for *Teacher-1***

| Answer | Category | 01 | 02 | 03 | 04 | 05 | 06 | 07 | 08 | 09 | 10 | 11 | 12 | 13 | 14 | 15 | 16 | 17 | 18 | 19 | 20 |
|---|---|---|---|---|---|---|---|---|---|---|---|---|---|---|---|---|---|---|---|---|---|
| 1 | 1 | 3 | 4 | 3 | 4 | 5 | 4 | 3 | 3 | 3 | 3 | 4 | 4 | 5 | 3 | 4 | 4 | 5 | 4 | 3 | 3 |
| 2 | 3 | 4 | 4 | 5 | 5 | 4 | 5 | 5 | 5 | 5 | 4 | 4 | 5 | 5 | 5 | 5 | 4 | 4 | 5 | 5 | 4 |
| 3 | 1 | 3 | 4 | 4 | 5 | 4 | 4 | 5 | 3 | 4 | 4 | 5 | 4 | 4 | 3 | 5 | 4 | 4 | 4 | 3 | 5 |
| 4 | 1 | 4 | 4 | 5 | 4 | 5 | 4 | 4 | 3 | 3 | 3 | 5 | 4 | 4 | 3 | 5 | 4 | 5 | 5 | 4 | 3 |
| 5 | 2 | 5 | 4 | 5 | 5 | 4 | 4 | 4 | 5 | 5 | 5 | 5 | 4 | 4 | 4 | 5 | 5 | 5 | 4 | 4 | 4 |
| 6 | 4 | 4 | 3 | 5 | 5 | 5 | 4 | 3 | 3 | 5 | 5 | 3 | 4 | 4 | 5 | 5 | 4 | 4 | 4 | 4 | 5 |
| 7 | 1 | 4 | 4 | 4 | 4 | 5 | 4 | 4 | 3 | 4 | 4 | 5 | 4 | 5 | 4 | 5 | 4 | 4 | 4 | 4 | 4 |
| 8 | 3 | 4 | 4 | 4 | 5 | 5 | 5 | 3 | 4 | 3 | 4 | 3 | 4 | 5 | 4 | 4 | 4 | 4 | 4 | 3 | 4 |
| 9 | 2 | 4 | 5 | 5 | 4 | 4 | 3 | 3 | 4 | 3 | 3 | 4 | 4 | 4 | 3 | 5 | 5 | 5 | 4 | 4 | 4 |
| 10 | 1 | 3 | 4 | 4 | 4 | 4 | 5 | 4 | 5 | 5 | 5 | 4 | 4 | 4 | 3 | 4 | 4 | 4 | 5 | 3 | 5 |

| Answer | Category | 01 | 02 | 03 | 04 | 05 | 06 | 07 | 08 | 09 | 10 | 11 | 12 | 13 | 14 | 15 | 16 | 17 | 18 | 19 | 20 |
|---|---|---|---|---|---|---|---|---|---|---|---|---|---|---|---|---|---|---|---|---|---|
| 11 | 3 | 4 | 4 | 3 | 4 | 5 | 3 | 4 | 3 | 5 | 4 | 4 | 4 | 4 | 4 | 5 | 4 | 4 | 4 | 3 | 5 |
| 12 | 4 | 4 | 5 | 5 | 4 | 5 | 5 | 4 | 4 | 5 | 5 | 5 | 5 | 5 | 5 | 5 | 4 | 5 | 5 | 4 | 5 |
| 13 | 1 | 5 | 5 | 5 | 5 | 4 | 5 | 4 | 5 | 5 | 5 | 5 | 5 | 4 | 3 | 5 | 4 | 5 | 5 | 5 | 4 |
| 14 | 3 | 4 | 4 | 4 | 5 | 5 | 4 | 3 | 5 | 4 | 3 | 5 | 4 | 4 | 4 | 3 | 4 | 4 | 4 | 4 | 4 |
| 15 | 2 | 4 | 3 | 3 | 4 | 5 | 4 | 4 | 4 | 4 | 4 | 5 | 4 | 4 | 4 | 5 | 4 | 4 | 4 | 4 | 4 |
| 16 | 3 | 4 | 4 | 3 | 4 | 5 | 5 | 4 | 5 | 3 | 5 | 5 | 3 | 4 | 4 | 5 | 4 | 4 | 5 | 4 | 3 |
| 17 | 4 | 3 | 4 | 5 | 5 | 5 | 5 | 5 | 4 | 5 | 5 | 4 | 4 | 5 | 4 | 5 | 5 | 5 | 4 | 4 | 4 |
| 18 | 2 | 4 | 4 | 5 | 5 | 4 | 4 | 4 | 3 | 5 | 4 | 4 | 4 | 4 | 4 | 5 | 4 | 4 | 5 | 4 | 4 |
| 19 | 3 | 4 | 4 | 5 | 5 | 4 | 4 | 5 | 5 | 5 | 4 | 4 | 4 | 4 | 4 | 5 | 4 | 4 | 4 | 4 | 4 |
| 20 | 2 | 5 | 5 | 4 | 5 | 4 | 3 | 3 | 3 | 4 | 5 | 4 | 4 | 5 | 4 | 4 | 4 | 5 | 5 | 5 | 5 |
| 21 | 1 | 4 | 5 | 4 | 4 | 5 | 3 | 4 | 3 | 4 | 3 | 5 | 3 | 5 | 4 | 5 | 4 | 5 | 4 | 5 | 5 |
| 22 | 4 | 3 | 4 | 5 | 5 | 5 | 5 | 4 | 4 | 3 | 5 | 4 | 4 | 5 | 5 | 5 | 4 | 5 | 4 | 5 | 4 |
| 23 | 2 | 4 | 4 | 5 | 4 | 4 | 4 | 4 | 5 | 4 | 5 | 4 | 4 | 4 | 5 | 4 | 5 | 4 | 5 | 5 | 4 |
| 24 | 4 | 5 | 4 | 5 | 5 | 5 | 5 | 5 | 5 | 3 | 5 | 5 | 5 | 4 | 5 | 5 | 4 | 5 | 5 | 5 | 3 |
| 25 | 1 | 4 | 4 | 3 | 4 | 5 | 3 | 5 | 3 | 3 | 4 | 4 | 3 | 4 | 3 | 5 | 4 | 4 | 5 | 4 | 4 |
| 26 | 3 | 4 | 4 | 4 | 5 | 5 | 4 | 5 | 5 | 3 | 5 | 3 | 4 | 4 | 4 | 5 | 5 | 4 | 5 | 4 | 4 |
| 27 | 2 | 5 | 4 | 4 | 5 | 5 | 4 | 4 | 4 | 3 | 4 | 4 | 4 | 5 | 4 | 5 | 4 | 5 | 4 | 4 | 4 |
| 28 | 1 | 3 | 5 | 5 | 5 | 5 | 3 | 4 | 5 | 4 | 3 | 4 | 4 | 5 | 3 | 4 | 4 | 3 | 3 | 4 | 4 |
| 29 | 4 | 5 | 4 | 4 | 5 | 5 | 3 | 4 | 4 | 3 | 5 | 3 | 5 | 5 | 4 | 4 | 4 | 4 | 4 | 4 | 4 |
| 30 | 3 | 4 | 4 | 4 | 5 | 4 | 5 | 5 | 5 | 4 | 5 | 4 | 4 | 5 | 4 | 5 | 4 | 4 | 5 | 4 | 4 |
| 31 | 4 | 4 | 4 | 3 | 5 | 5 | 4 | 5 | 4 | 4 | 4 | 3 | 4 | 4 | 4 | 5 | 3 | 4 | 5 | 4 | 4 |
| 32 | 2 | 4 | 5 | 4 | 4 | 4 | 5 | 4 | 4 | 4 | 3 | 4 | 4 | 4 | 4 | 5 | 4 | 5 | 4 | 4 | 4 |
| 33 | 1 | 4 | 4 | 5 | 4 | 4 | 3 | 4 | 3 | 3 | 5 | 4 | 3 | 4 | 3 | 4 | 4 | 4 | 4 | 5 | 5 |
| 34 | 4 | 4 | 4 | 5 | 5 | 5 | 5 | 4 | 3 | 4 | 5 | 4 | 4 | 5 | 4 | 5 | 4 | 5 | 4 | 5 | 5 |
| 35 | 3 | 3 | 5 | 4 | 5 | 4 | 4 | 5 | 5 | 3 | 4 | 4 | 5 | 4 | 5 | 5 | 5 | 4 | 4 | 4 | 5 |
| 36 | 3 | 4 | 4 | 5 | 5 | 5 | 5 | 5 | 5 | 3 | 5 | 4 | 4 | 5 | 4 | 5 | 4 | 4 | 3 | 4 | 4 |
| 37 | 4 | 5 | 4 | 5 | 4 | 4 | 4 | 4 | 3 | 5 | 5 | 4 | 4 | 4 | 5 | 5 | 4 | 5 | 4 | 4 | 3 |
| 38 | 2 | 4 | 4 | 4 | 4 | 4 | 4 | 4 | 3 | 4 | 5 | 5 | 4 | 4 | 4 | 5 | 5 | 5 | 5 | 4 | 4 |
| 39 | 4 | 3 | 4 | 5 | 5 | 4 | 5 | 4 | 4 | 4 | 3 | 5 | 4 | 4 | 4 | 5 | 4 | 5 | 4 | 4 | 5 |
| 40 | 1 | 4 | 4 | 5 | 5 | 4 | 4 | 4 | 4 | 4 | 5 | 5 | 4 | 5 | 5 | 5 | 3 | 4 | 4 | 4 | 5 |
| 41 | 4 | 4 | 4 | 5 | 5 | 4 | 5 | 5 | 3 | 4 | 5 | 5 | 4 | 4 | 5 | 4 | 5 | 4 | 4 | 4 | 4 |
| 42 | 2 | 4 | 3 | 3 | 4 | 5 | 3 | 3 | 3 | 3 | 5 | 5 | 3 | 4 | 3 | 5 | 4 | 4 | 3 | 3 | 4 |
| 43 | 1 | 4 | 4 | 5 | 5 | 5 | 3 | 5 | 4 | 4 | 4 | 5 | 4 | 4 | 5 | 5 | 5 | 5 | 5 | 5 | 5 |
| 44 | 3 | 5 | 5 | 5 | 5 | 4 | 5 | 5 | 5 | 5 | 5 | 5 | 4 | 5 | 4 | 5 | 4 | 5 | 4 | 4 | 4 |
| 45 | 4 | 4 | 4 | 4 | 5 | 4 | 3 | 4 | 4 | 4 | 5 | 5 | 5 | 5 | 4 | 4 | 5 | 4 | 4 | 4 | 5 |
| 46 | 2 | 5 | 4 | 5 | 5 | 5 | 5 | 5 | 5 | 5 | 4 | 5 | 5 | 5 | 4 | 5 | 4 | 5 | 5 | 4 | 5 |
| 47 | 2 | 5 | 5 | 5 | 4 | 5 | 4 | 4 | 5 | 4 | 5 | 5 | 5 | 5 | 4 | 4 | 3 | 5 | 5 | 4 | 4 |
| 48 | 3 | 5 | 5 | 5 | 5 | 5 | 5 | 4 | 5 | 5 | 5 | 5 | 5 | 5 | 5 | 4 | 4 | 5 | 5 | 4 | 5 |
| 49 | 2 | 4 | 5 | 5 | 5 | 5 | 5 | 5 | 5 | 5 | 5 | 4 | 4 | 4 | 4 | 4 | 5 | 4 | 4 | 4 | 5 |
| 50 | 2 | 5 | 5 | 4 | 5 | 5 | 4 | 5 | 5 | 5 | 5 | 5 | 4 | 5 | 4 | 4 | 4 | 5 | 5 | 5 | 5 |
| 51 | 2 | 5 | 5 | 5 | 4 | 5 | 5 | 4 | 5 | 4 | 5 | 5 | 4 | 5 | 4 | 5 | 5 | 5 | 5 | 5 | 4 |
| 52 | 2 | 5 | 5 | 4 | 5 | 4 | 5 | 4 | 5 | 4 | 5 | 5 | 5 | 5 | 5 | 5 | 5 | 5 | 4 | 5 | 4 |
| 53 | 2 | 5 | 4 | 5 | 5 | 5 | 4 | 4 | 5 | 4 | 4 | 4 | 4 | 5 | 5 | 5 | 5 | 5 | 5 | 5 | 4 |
| 54 | 2 | 5 | 5 | 5 | 4 | 5 | 5 | 5 | 5 | 5 | 5 | 5 | 5 | 4 | 5 | 5 | 5 | 4 | 5 | 5 | 5 |
| 55 | 3 | 4 | 4 | 4 | 5 | 4 | 5 | 5 | 5 | 4 | 5 | 5 | 4 | 5 | 4 | 5 | 4 | 5 | 4 | 5 | 5 |
| 56 | 2 | 5 | 5 | 4 | 5 | 5 | 5 | 5 | 5 | 4 | 4 | 4 | 5 | 5 | 5 | 4 | 5 | 5 | 5 | 5 | 5 |
| 57 | 4 | 5 | 5 | 5 | 4 | 5 | 4 | 5 | 5 | 4 | 5 | 5 | 4 | 5 | 4 | 5 | 4 | 5 | 5 | 5 | 5 |
| 58 | 2 | 5 | 5 | 5 | 4 | 5 | 5 | 5 | 5 | 4 | 4 | 5 | 5 | 5 | 4 | 4 | 5 | 4 | 5 | 5 | 4 |

Anale. Seria Informatică. Vol. VII fasc. 1 – 2009
Annals. Computer Science Series. 7th Tome 1st Fasc. – 2009## Results

So, first processing step is to order the answers according to the competence (Table 2) and then to obtain some statistical parameters ad minimum and maximum, medium value, standard deviation, number of different results (Table 3).

**Table 2: Questionnaires' results for Teacher-1 ordered by competencies**

| Answer | Category | 01 | 02 | 03 | 04 | 05 | 06 | 07 | 08 | 09 | 10 | 11 | 12 | 13 | 14 | 15 | 16 | 17 | 18 | 19 | 20 |
|---|---|---|---|---|---|---|---|---|---|---|---|---|---|---|---|---|---|---|---|---|---|
| 1  | 1 | 3 | 4 | 3 | 4 | 5 | 4 | 3 | 3 | 3 | 3 | 4 | 4 | 5 | 3 | 4 | 4 | 5 | 4 | 3 | 3 |
| 3  | 1 | 3 | 4 | 4 | 5 | 4 | 4 | 5 | 3 | 4 | 4 | 5 | 4 | 4 | 3 | 5 | 4 | 4 | 4 | 3 | 5 |
| 4  | 1 | 4 | 4 | 5 | 4 | 5 | 4 | 4 | 3 | 3 | 3 | 5 | 4 | 4 | 3 | 5 | 4 | 5 | 5 | 4 | 3 |
| 7  | 1 | 4 | 4 | 4 | 4 | 5 | 4 | 4 | 3 | 4 | 4 | 5 | 4 | 5 | 4 | 5 | 4 | 4 | 4 | 4 | 4 |
| 10 | 1 | 3 | 4 | 4 | 4 | 4 | 5 | 4 | 5 | 5 | 5 | 4 | 4 | 4 | 3 | 4 | 4 | 4 | 5 | 3 | 5 |
| 13 | 1 | 5 | 5 | 5 | 5 | 4 | 5 | 4 | 5 | 5 | 5 | 5 | 5 | 4 | 3 | 5 | 4 | 5 | 5 | 5 | 4 |
| 21 | 1 | 4 | 5 | 4 | 4 | 5 | 3 | 4 | 3 | 4 | 3 | 5 | 3 | 5 | 4 | 5 | 4 | 5 | 4 | 5 | 5 |
| 25 | 1 | 4 | 4 | 3 | 4 | 5 | 3 | 5 | 3 | 3 | 4 | 4 | 3 | 4 | 3 | 5 | 4 | 4 | 5 | 4 | 4 |
| 28 | 1 | 3 | 5 | 5 | 5 | 5 | 3 | 4 | 5 | 4 | 3 | 4 | 4 | 5 | 3 | 4 | 4 | 3 | 3 | 4 | 4 |
| 33 | 1 | 4 | 4 | 5 | 4 | 4 | 3 | 4 | 3 | 3 | 5 | 4 | 3 | 4 | 3 | 4 | 4 | 4 | 4 | 5 | 5 |
| 40 | 1 | 4 | 4 | 5 | 5 | 4 | 4 | 4 | 4 | 4 | 5 | 5 | 4 | 5 | 5 | 5 | 3 | 4 | 4 | 4 | 5 |
| 43 | 1 | 4 | 4 | 5 | 5 | 5 | 3 | 5 | 4 | 4 | 5 | 4 | 4 | 4 | 5 | 5 | 5 | 5 | 5 | 5 | 5 |
| 5  | 2 | 5 | 4 | 5 | 5 | 4 | 4 | 4 | 5 | 5 | 5 | 5 | 4 | 4 | 4 | 5 | 5 | 5 | 4 | 4 | 4 |
| 9  | 2 | 4 | 5 | 5 | 4 | 4 | 3 | 3 | 4 | 3 | 3 | 4 | 4 | 4 | 3 | 5 | 5 | 5 | 4 | 4 | 4 |
| 15 | 2 | 4 | 3 | 3 | 4 | 5 | 4 | 4 | 4 | 4 | 4 | 5 | 4 | 4 | 4 | 5 | 4 | 4 | 4 | 4 | 4 |
| 18 | 2 | 4 | 4 | 5 | 5 | 4 | 4 | 4 | 3 | 5 | 4 | 4 | 4 | 4 | 4 | 5 | 4 | 4 | 5 | 4 | 4 |
| 20 | 2 | 5 | 5 | 4 | 5 | 4 | 3 | 3 | 3 | 4 | 5 | 4 | 4 | 5 | 4 | 4 | 4 | 5 | 5 | 5 | 5 |
| 23 | 2 | 4 | 4 | 5 | 4 | 4 | 4 | 4 | 5 | 4 | 4 | 5 | 4 | 4 | 4 | 5 | 4 | 5 | 5 | 5 | 4 |
| 27 | 2 | 5 | 4 | 4 | 5 | 5 | 4 | 4 | 4 | 3 | 4 | 4 | 4 | 5 | 4 | 5 | 4 | 5 | 4 | 4 | 4 |
| 32 | 2 | 4 | 5 | 4 | 4 | 4 | 5 | 4 | 4 | 4 | 3 | 4 | 4 | 4 | 4 | 5 | 4 | 5 | 4 | 4 | 4 |
| 38 | 2 | 4 | 4 | 4 | 4 | 4 | 4 | 4 | 4 | 3 | 4 | 5 | 5 | 4 | 4 | 5 | 5 | 5 | 5 | 4 | 4 |
| 42 | 2 | 4 | 3 | 3 | 4 | 5 | 3 | 3 | 3 | 3 | 5 | 5 | 3 | 4 | 3 | 5 | 4 | 4 | 3 | 3 | 4 |
| 46 | 2 | 5 | 4 | 5 | 5 | 5 | 5 | 5 | 5 | 5 | 4 | 5 | 5 | 5 | 4 | 5 | 4 | 5 | 5 | 4 | 5 |
| 47 | 2 | 5 | 5 | 5 | 4 | 5 | 4 | 4 | 5 | 4 | 5 | 5 | 5 | 5 | 4 | 4 | 3 | 5 | 5 | 4 | 4 |
| 49 | 2 | 4 | 5 | 5 | 5 | 5 | 5 | 5 | 5 | 5 | 5 | 4 | 4 | 4 | 4 | 4 | 4 | 5 | 4 | 4 | 5 |
| 50 | 2 | 5 | 5 | 4 | 5 | 5 | 4 | 5 | 5 | 5 | 5 | 5 | 4 | 4 | 5 | 4 | 4 | 5 | 5 | 5 | 5 |
| 51 | 2 | 5 | 5 | 5 | 4 | 5 | 5 | 4 | 5 | 4 | 5 | 5 | 4 | 5 | 4 | 5 | 5 | 5 | 5 | 5 | 4 |
| 52 | 2 | 5 | 5 | 4 | 5 | 4 | 5 | 4 | 5 | 4 | 5 | 5 | 5 | 5 | 5 | 5 | 5 | 5 | 4 | 5 | 4 |
| 53 | 2 | 5 | 4 | 5 | 5 | 5 | 4 | 4 | 5 | 4 | 4 | 4 | 4 | 5 | 5 | 5 | 5 | 5 | 5 | 5 | 4 |
| 54 | 2 | 5 | 5 | 5 | 4 | 5 | 5 | 5 | 5 | 5 | 5 | 5 | 5 | 4 | 5 | 5 | 5 | 4 | 5 | 5 | 5 |
| 56 | 2 | 5 | 5 | 4 | 5 | 5 | 5 | 5 | 5 | 5 | 4 | 4 | 4 | 5 | 5 | 5 | 4 | 5 | 5 | 5 | 5 |
| 58 | 2 | 5 | 5 | 5 | 4 | 5 | 5 | 5 | 5 | 4 | 4 | 5 | 5 | 5 | 4 | 4 | 5 | 4 | 5 | 4 | 4 |
| 2  | 3 | 4 | 4 | 5 | 5 | 4 | 5 | 5 | 5 | 5 | 4 | 4 | 5 | 5 | 5 | 5 | 4 | 4 | 5 | 5 | 4 |
| 8  | 3 | 4 | 4 | 4 | 5 | 5 | 5 | 3 | 4 | 3 | 4 | 3 | 4 | 5 | 4 | 4 | 4 | 4 | 4 | 3 | 4 |
| 11 | 3 | 4 | 4 | 3 | 4 | 5 | 3 | 4 | 3 | 5 | 4 | 4 | 4 | 4 | 4 | 5 | 4 | 4 | 4 | 3 | 5 |
| 14 | 3 | 4 | 4 | 4 | 5 | 5 | 4 | 3 | 5 | 4 | 3 | 5 | 4 | 4 | 4 | 3 | 4 | 4 | 4 | 4 | 4 |
| 16 | 3 | 4 | 4 | 3 | 4 | 5 | 5 | 4 | 5 | 3 | 5 | 5 | 3 | 4 | 4 | 5 | 4 | 4 | 5 | 4 | 3 |
| 19 | 3 | 4 | 4 | 5 | 5 | 4 | 4 | 5 | 5 | 5 | 4 | 4 | 4 | 4 | 4 | 5 | 4 | 4 | 4 | 4 | 4 |
| 26 | 3 | 4 | 4 | 4 | 5 | 5 | 4 | 5 | 5 | 3 | 5 | 3 | 4 | 4 | 4 | 5 | 5 | 4 | 5 | 4 | 4 |

184

| Answer | Category | 01 | 02 | 03 | 04 | 05 | 06 | 07 | 08 | 09 | 10 | 11 | 12 | 13 | 14 | 15 | 16 | 17 | 18 | 19 | 20 |
|---|---|---|---|---|---|---|---|---|---|---|---|---|---|---|---|---|---|---|---|---|---|
| 30 | 3 | 4 | 4 | 4 | 5 | 4 | 5 | 5 | 5 | 4 | 5 | 4 | 4 | 5 | 4 | 5 | 4 | 4 | 5 | 4 | 4 |
| 35 | 3 | 3 | 5 | 4 | 5 | 4 | 4 | 5 | 5 | 3 | 4 | 4 | 5 | 4 | 5 | 5 | 5 | 4 | 4 | 4 | 5 |
| 36 | 3 | 4 | 4 | 5 | 5 | 5 | 5 | 5 | 5 | 5 | 3 | 5 | 4 | 4 | 5 | 4 | 4 | 4 | 3 | 4 | 4 |
| 44 | 3 | 5 | 5 | 5 | 5 | 4 | 5 | 5 | 5 | 5 | 5 | 5 | 4 | 5 | 4 | 5 | 4 | 5 | 4 | 4 | 4 |
| 48 | 3 | 5 | 5 | 5 | 5 | 5 | 5 | 4 | 5 | 5 | 5 | 5 | 5 | 5 | 4 | 4 | 5 | 5 | 4 | 4 | 5 |
| 55 | 3 | 4 | 4 | 4 | 5 | 4 | 5 | 5 | 5 | 5 | 4 | 4 | 5 | 5 | 4 | 5 | 4 | 5 | 4 | 5 | 5 |
| 6 | 4 | 4 | 3 | 5 | 5 | 5 | 4 | 3 | 3 | 5 | 5 | 3 | 4 | 4 | 5 | 5 | 4 | 4 | 4 | 4 | 5 |
| 12 | 4 | 4 | 5 | 5 | 4 | 5 | 5 | 4 | 4 | 5 | 5 | 5 | 5 | 5 | 5 | 5 | 4 | 5 | 5 | 4 | 5 |
| 17 | 4 | 3 | 4 | 5 | 5 | 5 | 5 | 5 | 4 | 5 | 5 | 4 | 4 | 5 | 4 | 5 | 5 | 5 | 4 | 4 | 4 |
| 22 | 4 | 3 | 4 | 5 | 5 | 5 | 5 | 4 | 4 | 3 | 5 | 4 | 4 | 5 | 5 | 5 | 4 | 5 | 4 | 5 | 4 |
| 24 | 4 | 5 | 4 | 5 | 5 | 5 | 5 | 5 | 5 | 3 | 5 | 5 | 5 | 4 | 5 | 5 | 4 | 5 | 5 | 5 | 3 |
| 29 | 4 | 5 | 4 | 4 | 5 | 5 | 3 | 4 | 4 | 3 | 5 | 3 | 5 | 5 | 4 | 4 | 4 | 4 | 4 | 4 | 4 |
| 31 | 4 | 4 | 4 | 3 | 5 | 5 | 4 | 5 | 4 | 4 | 4 | 3 | 4 | 4 | 4 | 5 | 3 | 4 | 5 | 4 | 4 |
| 34 | 4 | 4 | 4 | 5 | 5 | 5 | 5 | 4 | 3 | 4 | 5 | 4 | 4 | 5 | 4 | 5 | 4 | 5 | 4 | 5 | 5 |
| 37 | 4 | 5 | 4 | 5 | 4 | 4 | 4 | 4 | 3 | 5 | 5 | 4 | 4 | 4 | 5 | 5 | 4 | 5 | 4 | 4 | 3 |
| 39 | 4 | 3 | 4 | 5 | 5 | 4 | 5 | 4 | 4 | 4 | 3 | 5 | 4 | 4 | 4 | 5 | 4 | 5 | 4 | 4 | 5 |
| 41 | 4 | 4 | 4 | 5 | 5 | 4 | 5 | 5 | 3 | 4 | 5 | 5 | 4 | 4 | 4 | 5 | 4 | 5 | 4 | 4 | 4 |
| 45 | 4 | 4 | 4 | 4 | 5 | 4 | 3 | 4 | 4 | 4 | 5 | 5 | 5 | 5 | 4 | 4 | 4 | 5 | 4 | 4 | 5 |
| 57 | 4 | 5 | 5 | 5 | 4 | 5 | 4 | 5 | 5 | 4 | 5 | 5 | 4 | 5 | 4 | 5 | 4 | 5 | 5 | 5 | 5 |

**Table 3: Questionnaires' results for Teacher-1 – statistical results**

| Answer | Category | Min | Max | Medium | Std.dev. | No.1 | No.2 | No.3 | No.4 | No.5 |
|---|---|---|---|---|---|---|---|---|---|---|
| 1 | 1 | 3 | 5 | 3,70 | 0,73270 | | | 9 | 8 | 8 |
| 3 | 1 | 3 | 5 | 4,05 | 0,68633 | | | 4 | 11 | 11 |
| 4 | 1 | 3 | 5 | 4,05 | 0,75915 | | | 5 | 9 | 9 |
| 7 | 1 | 3 | 5 | 4,15 | 0,48936 | | | 1 | 15 | 15 |
| 10 | 1 | 3 | 5 | 4,15 | 0,67082 | | | 3 | 11 | 11 |
| 13 | 1 | 3 | 5 | 4,65 | 0,58714 | | | 1 | 5 | 5 |
| 21 | 1 | 3 | 5 | 4,20 | 0,76777 | | | 4 | 8 | 8 |
| 25 | 1 | 3 | 5 | 3,90 | 0,71818 | | | 6 | 10 | 10 |
| 28 | 1 | 3 | 5 | 4,00 | 0,79472 | | | 6 | 8 | 8 |
| 33 | 1 | 3 | 5 | 3,95 | 0,68633 | | | 5 | 11 | 11 |
| 40 | 1 | 3 | 5 | 4,35 | 0,58714 | | | 1 | 11 | 11 |
| 43 | 1 | 3 | 5 | 4,50 | 0,60698 | | | 1 | 8 | 8 |
| 5 | 2 | 4 | 5 | 4,50 | 0,51299 | | | | 10 | 10 |
| 9 | 2 | 3 | 5 | 4,00 | 0,72548 | | | 5 | 10 | 10 |
| 15 | 2 | 3 | 5 | 4,05 | 0,51042 | | | 2 | 15 | 15 |
| 18 | 2 | 3 | 5 | 4,20 | 0,52315 | | | 1 | 14 | 14 |
| 20 | 2 | 3 | 5 | 4,30 | 0,73270 | | | 3 | 8 | 8 |
| 23 | 2 | 4 | 5 | 4,35 | 0,48936 | | | | 13 | 13 |
| 27 | 2 | 3 | 5 | 4,25 | 0,55012 | | | 1 | 13 | 13 |
| 32 | 2 | 3 | 5 | 4,15 | 0,48936 | | | 1 | 15 | 15 |
| 38 | 2 | 3 | 5 | 4,25 | 0,55012 | | | 1 | 13 | 13 |
| 42 | 2 | 3 | 5 | 3,70 | 0,80131 | | | 10 | 6 | 6 |
| 46 | 2 | 4 | 5 | 4,75 | 0,44426 | | | | 5 | 5 |
| 47 | 2 | 3 | 5 | 4,50 | 0,60698 | | | 1 | 8 | 8 |
| 49 | 2 | 4 | 5 | 4,55 | 0,51042 | | | | 9 | 9 |
| 50 | 2 | 4 | 5 | 4,70 | 0,47016 | | | | 6 | 6 |
| 51 | 2 | 4 | 5 | 4,70 | 0,47016 | | | | 6 | 6 |

185

Anale. Seria Informatică. Vol. VII fasc. 1 – 2009
Annals. Computer Science Series. 7th Tome 1st Fasc. – 2009| | | | | | | | | | | |
|---|---|---|---|---|---|---|---|---|---|---|
| 52 | 2 | 4 | 5 | 4,70 | 0,47016 | | | | 6 | 6 |
| 53 | 2 | 4 | 5 | 4,55 | 0,51042 | | | | 9 | 9 |
| 54 | 2 | 4 | 5 | 4,85 | 0,36635 | | | | 3 | 3 |
| 56 | 2 | 4 | 5 | 4,75 | 0,44426 | | | | 5 | 5 |
| 58 | 2 | 4 | 5 | 4,65 | 0,48936 | | | | 7 | 7 |
| 2 | 3 | 4 | 5 | 4,60 | 0,50262 | | | | 8 | 8 |
| 8 | 3 | 3 | 5 | 4,00 | 0,64889 | | | 4 | 12 | 12 |
| 11 | 3 | 3 | 5 | 4,00 | 0,64889 | | | 4 | 12 | 12 |
| 14 | 3 | 3 | 5 | 4,05 | 0,60481 | | | 3 | 13 | 13 |
| 16 | 3 | 3 | 5 | 4,15 | 0,74516 | | | 4 | 9 | 9 |
| 19 | 3 | 4 | 5 | 4,30 | 0,47016 | | | | 14 | 14 |
| 26 | 3 | 3 | 5 | 4,30 | 0,65695 | | | 2 | 10 | 10 |
| 30 | 3 | 4 | 5 | 4,40 | 0,50262 | | | | 12 | 12 |
| 35 | 3 | 3 | 5 | 4,35 | 0,67082 | | | 2 | 9 | 9 |
| 36 | 3 | 3 | 5 | 4,35 | 0,67082 | | | 2 | 9 | 9 |
| 44 | 3 | 4 | 5 | 4,65 | 0,48936 | | | | 7 | 7 |
| 48 | 3 | 4 | 5 | 4,80 | 0,41039 | | | | 4 | 4 |
| 55 | 3 | 4 | 5 | 4,55 | 0,51042 | | | | 9 | 9 |
| 6 | 4 | 3 | 5 | 4,20 | 0,76777 | | | 4 | 8 | 8 |
| 12 | 4 | 4 | 5 | 4,70 | 0,47016 | | | | 6 | 6 |
| 17 | 4 | 3 | 5 | 4,50 | 0,60698 | | | 1 | 8 | 8 |
| 22 | 4 | 3 | 5 | 4,40 | 0,68056 | | | 2 | 8 | 8 |
| 24 | 4 | 3 | 5 | 4,65 | 0,67082 | | | 2 | 3 | 3 |
| 29 | 4 | 3 | 5 | 4,15 | 0,67082 | | | 3 | 11 | 11 |
| 31 | 4 | 3 | 5 | 4,10 | 0,64072 | | | 3 | 12 | 12 |
| 34 | 4 | 3 | 5 | 4,45 | 0,60481 | | | 1 | 9 | 9 |
| 37 | 4 | 3 | 5 | 4,25 | 0,63867 | | | 2 | 11 | 11 |
| 39 | 4 | 3 | 5 | 4,25 | 0,63867 | | | 2 | 11 | 11 |
| 41 | 4 | 3 | 5 | 4,35 | 0,58714 | | | 1 | 11 | 11 |
| 45 | 4 | 3 | 5 | 4,30 | 0,57124 | | | 1 | 12 | 12 |
| 57 | 4 | 4 | 5 | 4,70 | 0,47016 | | | | 6 | 6 |

Next, data are aggregated depending on the competence and, again, the statistical parameters as minimum and maximum, medium value, standard deviation, number of different results are determined; further, the general results are processed (Table 4):

**Table 4: Questionnaires' results for Teacher-1 – aggregated results**

| Category | Min | Max | Medium | Std.dev. | No.1 | No.2 | No.3 | No.4 | No.5 |
|---|---|---|---|---|---|---|---|---|---|
| 1 | 3 | 5 | 4,14 | 0,70995 | | | 46 | 115 | 79 |
| 2 | 3 | 5 | 4,42 | 0,60821 | | | 25 | 181 | 194 |
| 3 | 3 | 5 | 4,35 | 0,62399 | | | 21 | 128 | 111 |
| 4 | 3 | 5 | 4,38 | 0,63835 | | | 22 | 116 | 122 |
| **TOTAL** | **3** | **5** | **4,34** | **0.64857** | | | **114** | **540** | **506** |

The program also makes a graphical representation for all these information as depicted in Figure 1 and Figure 2; the Internet page containing the results of the processing is depicted in Figure 3.

186

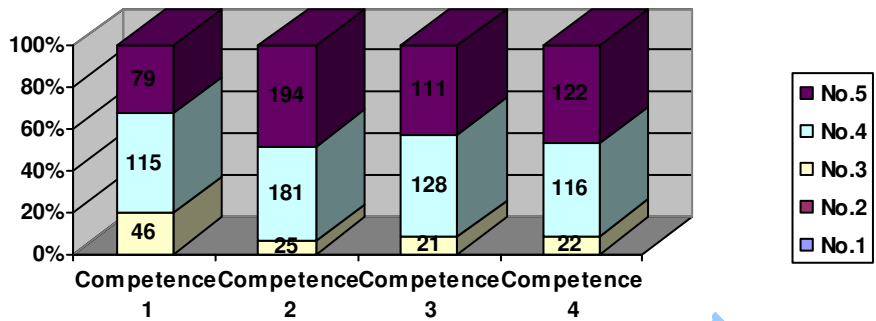

**Figure 1: Number of marks depending on competencies**

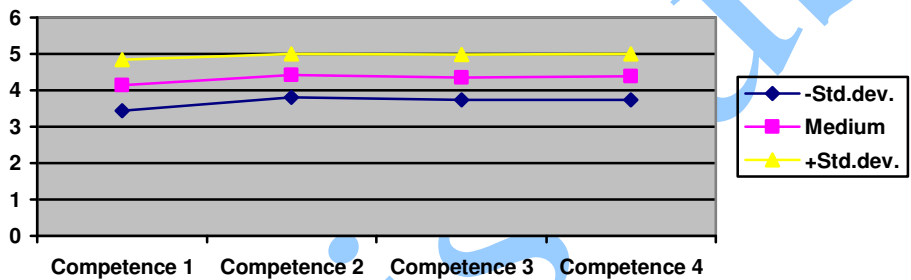

**Figure 2: Mark intervals depending on competencies**

Statistic results for:

Teacher-1

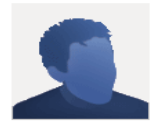

| Competence | Min | Max | Medium | Std.dev. | No.1 | No.2 | No.3 | No.4 | No.5 |
|---|---|---|---|---|---|---|---|---|---|
| 1 | 3 | 5 | 4.14 | 0.70995 | 16 | 23 | 46 | 115 | 79 |
| 2 | 3 | 5 | 4.42 | 0.60821 | 24 | 35 | 25 | 181 | 194 |
| 3 | 3 | 5 | 4.35 | 0.62399 | 14 | 43 | 21 | 128 | 111 |
| 4 | 3 | 5 | 4.38 | 0.63835 | 55 | 12 | 22 | 116 | 122 |

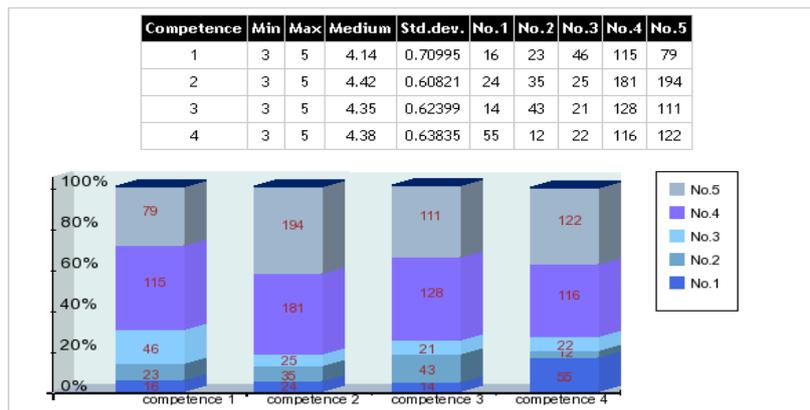

**Figure 3: On-line applications' results**

**Conclusion**

This application provides to the administrator the advantage of automatic calculation of the medium, minimum and maximum marks values, eliminating the human intervention in computing and graphical representation of the results.